# Search for Hidden Photon Dark Matter Using a Multi-Cathode Counter

A. Kopylov*, I. Orekhov, V. Petukhov

Institute for nuclear research RAS, Moscow 117312, prospect 60-letiya Octyabrya 7a

beril@inr.ru

**Abstract**. A search for hidden-photon (HP) dark matter using a multi-cathode counter is reported. The technique based on counting of single electrons emitted from outer cathode of the proportional counter by hidden-photons was used. The apparatus and the calibration of the counter by ultraviolet lamp are described. It is shown that this technique attains a maximum sensitivity in the energy range of Vacuum Ultraviolet. From the results of measurements we set an upper limit on the photon-HP mixing parameter χ. A further progress of using a multi-cathode technique is discussed.

1. **Introduction.**

Hidden photons (HPs) were proposed by L. B. Okun [1] in 1982 as a possible modification of electrodynamics. The hypothetical HPs are also interesting as one of the alternatives for Cold Dark Matter (CDM). A novel approach to register HPs using a dish antenna was proposed [2] and the novel technique has been realized [3]. The range of sensitivity in this experiment 3.1±1.2 eV was limited by the coefficient of reflection of light from a mirror and by spectral sensitivity of photomultiplier used. By higher energies the light gets absorbed by the surface of the mirror and the sensitivity deteriorates. When an ultraviolet light gets absorbed by metal surface the electrons are emitted. So if to register electrons emitted from the surface of a metallic absorber one can extend the sensitivity of the method of a dish antenna in the range of Vacuum Ultraviolet and further. If dark matter is totally composed of HPs, then the power collected by an antenna (here, by a cathode of the counter).

$$P = 2\alpha^2 \chi^2 \rho_{CDM} A_{cath} \qquad (1)$$

where $\alpha^2 = \cos\theta$ and $\theta$ is the angle between the direction of HP vector and the surface of the cathode, $\alpha^2 = 2/3$ if the HP vector is isotropic, $\rho_{CDM} \approx 0.3$ GeV/cm$^3$ is the energy density of CDM, which is assumed to be equal to the energy density of the HPs, $A_{cath}$ is the area of the cathode of the counter and χ is a dimensionless parameter quantifying the kinetic mixing [2]. If this power is converted into single electrons emitted from the cathode of the counter, then

$$P = m_{\gamma'} \cdot R_{MCC} / \eta \qquad (2)$$

where: $m_{\gamma'}$ is the mass (energy) of an HP, $\eta$ is the quantum efficiency for a photon with energy $m_{\gamma'}$ to yield a single electron from the surface of the metal, and $R_{MCC}$ is the rate of single-electrons emitted from the cathode which is presumed here as being from HP. Thus, by combining (1) and (2) we obtain:

$$\chi_{sens} = 2.9 \cdot 10^{-12} \left( \frac{R_{MCC}}{\eta \cdot 1Hz} \right)^{1/2} \left( \frac{m_{\gamma'}}{1eV} \right)^{1/2} \left( \frac{0.3 GeV/cm^3}{\rho_{CDM}} \right)^{1/2} \left( \frac{1m^2}{A_{MCC}} \right)^{1/2} \left( \frac{\sqrt{2/3}}{\alpha} \right) \quad (3)$$

## 2. Experimental apparatus.

A schematic and a design of MCC were presented in [4, 5]. Here they are shown at figure 1. To reduce the background from the surrounding γ-radiation the counter has been placed in a special cabinet having as a passive shield the steel slabs in total 30 cm from all sides. The measurements performed inside and outside of the steel cabinet have shown that the background of single-electron events from external γ-radiation has been attenuated till the level of about 0.2% in comparison with the measured rate inside a shield. No single-electron events were observed in coincidence with muons passing through detector.

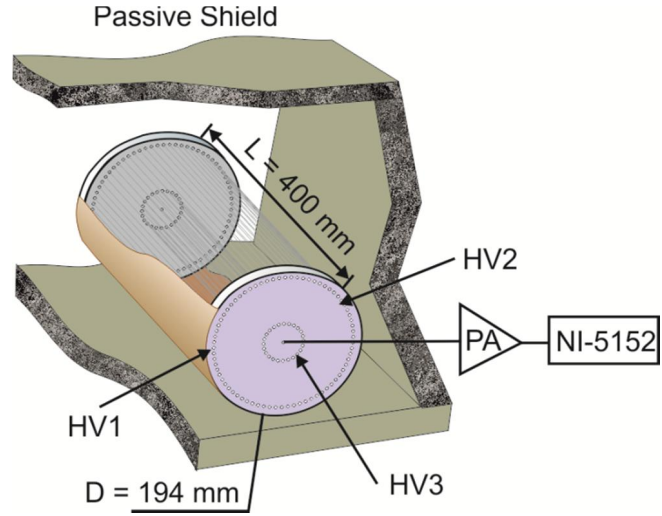

Figure 1. Schematic of MCC.

The counter is encapsulated in a hermetic stainless steel cylinder. All electrical connections are made through vacuum seals. It was filled with a mixture of Ar + 10% $CH_4$ at a pressure of about 0.1 MPa. The counter has several cathodes with negative high voltage applied to each cathode. First cathode is made of metal and is emitting individual electrons upon the absorption of UV light during calibration or as a result of hypothetical conversion of HPs. At a distance of 8 mm from the first cathode a second one is placed made of array of nichrome wires 50 μm in diameter interspaced by 8 mm. Different high voltages are applied to this cathode: in configuration 1 the potential of second cathode do not prevent electrons to move towards central counter while in configuration 2 the potential applied to second cathode acts as a barrier, so that electrons are scattered back towards first cathode. Cathode 3 serves as a cathode of central counter with high (~$10^5$) gas amplification which enables to detect single-electrons. Central wire of the counter is made of gold-plated W-Re alloy 25 μm in diameter. The signal from central wire is fed to the input of charge-sensitive preamplifier and then to the input of 8-bit digitizer. All data collected during measurements are stored on a disk and are analyzed off-line. The effect is evaluated by the difference of the count rates in first and second configuration. It was presumed that in first configuration we measure the count rate $R_1$ of single-electrons emitted: (1) from the first cathode, (2) from the surfaces at the ends of the counter, (3) from the multiple wires inside the counter while in the second configuration we measure the rate of single-electrons only from second and third components. Electrons emitted from first cathode are scattered back in second configuration and do not contribute to the total count rate $R_2$. So difference $R_{MCC} = R_1 - R_2$ of the count rates in first and second configurations should give the net effect from single-electrons emitted from an external cathode. To find the efficiency of the counter for single-electron measurements the calibration has been performed by UV mercury vapor lamp through a quartz window in the side of the counter [6]. The data treatment was performed in off-line. For the selection of "true" events a selection has been performed in space of three parameters: amplitude of the pulse, the duration of the leading edge of the pulse and a parameter β which describes prehistory of the event and is proportional to a first derivative of a baseline, approximated by a straight line during 50 μs before leading edge of the pulse. The efficiency was estimated as the probability for the pulse to belong to ROI box of this 3-parameter space. It was found to be (88 ± 6) %. To reduce the influence of the noise on counting only intervals with a baseline deviation from zero not more than 5 mV were taken into account with a proper correction for a live time of counting which was found to be about 54%.

**3. The results obtained.**

The measurements have been performed during 78 days at 26°C, 31°C and 36°C. During this time 15 TB of information has been obtained on the shapes of pulses of different amplitudes. Then the pulses were selected by amplitude, duration of a leading edge and prehistory of the event. Figure 3 shows the region of interest (ROI) used for selection of "true" pulses and distribution of the events for all three sets of measurements. The selection of ROI has been performed from data obtained by calibration of the counter by means of UV Mercury vapor lamp [6]. As the "true" pulses were accepted only the pulses having amplitudes from 3 till 30 mV, duration of the front edge from 2 till 25 μs and a parameter β which is dependent from prehistory of the event from -0.1 till 0.1 which excludes events with strong distortion of the baseline prior the front edge of the pulse. One can see that all three sets of measurements have similar distribution of pulses which proves the stability of the measurements. The efficiency of counting as a probability of the event to belong to ROI was estimated from calibration to be (88 ± 6) %.

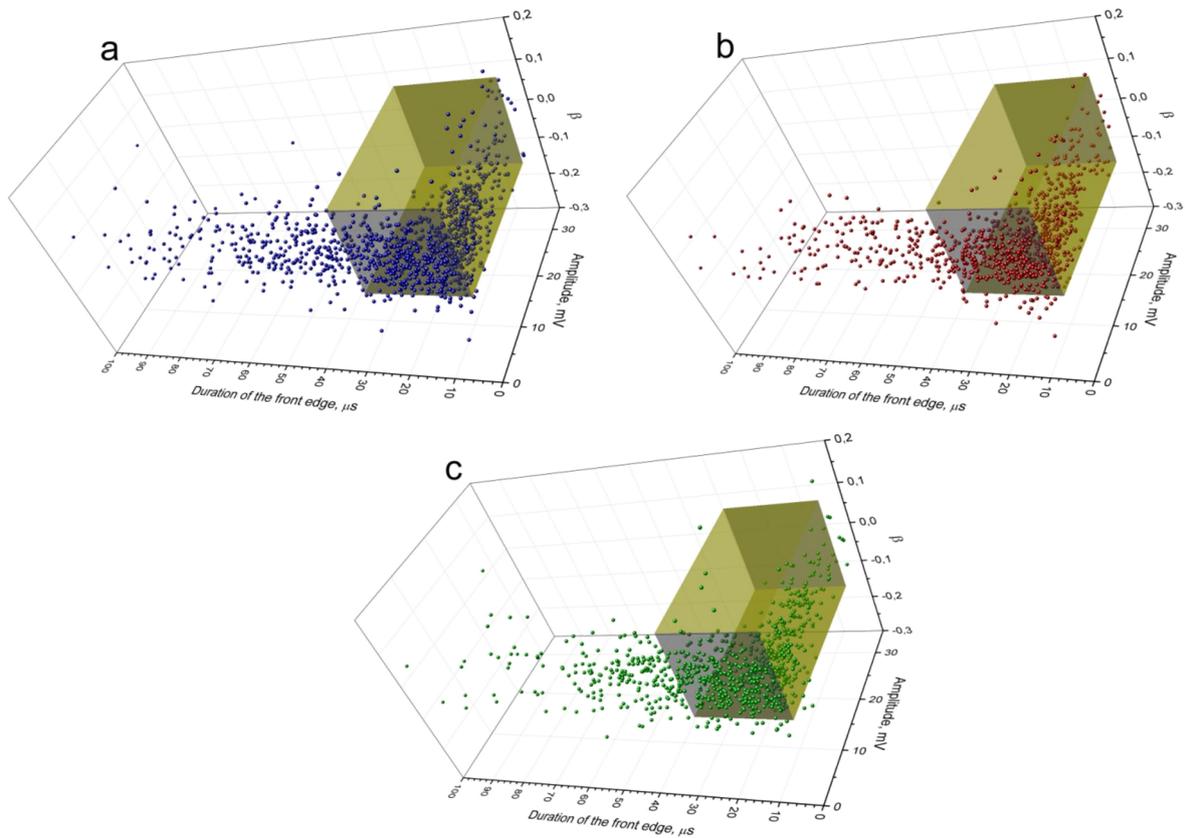

Figure 2. The distribution of the events in 3D space of parameters: amplitude of the pulse, duration of the front edge of the pulse and parameter β which is dependent upon the prehistory of the event. The distributions are presented for measurements at different temperatures: $26^{\circ}C$(a), $31^{\circ}C$(b), $36^{\circ}C$(c). The shaded area is the ROI used for the data analysis.

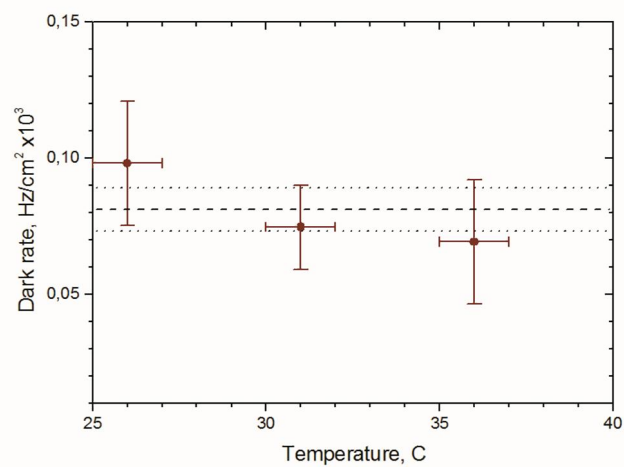

Figure 3. The distribution of events in sets $26^{\circ}C$, $31^{\circ}C$, $36^{\circ}C$. Dashed line – the average value obtained for all measurements, by point line are depicted levels ±1σ from average value.

One can see that all three sets have close distributions. After selection of "true" pulses we obtained for $r_{MCC} = R_{MCC}/A_{cath}$: $(0.98 \pm 0.22) \cdot 10^{-4}$ $Hz/cm^2$, $(0.75 \pm 0.15) \cdot 10^{-4}$ $Hz/cm^2$ and $(0.69 \pm 0.23) \cdot 10^{-4}$

Hz/cm² for these temperatures correspondingly. The fact that the count rate was not revealing a clear increase with temperature, as one can see from figure 3, can be taken as evidence that there was negligible contribution of thermal emission. As an average value it was obtained: $r_{MCC}$ = (0.81 ± 0.08)·$10^{-4}$ Hz/cm².

The rates measured by the same technique but using a new counter with an aluminum cathode and guarding rings [6] was shown to be much lower as one can see at figure 4. By averaging the points presented at Fig.4 we obtained: $r_{MCC}$ = (0.8 ± 0.25)·$10^{-5}$ Hz/cm². This is the lowest result obtained by the present time.

A summary of the results is presented below at figure 5. All limits presented at this figure are at confidence level 95%. The limits were obtained in the assumption that quantum efficiencies η to emit electron for a photon and for a HP are equal. The limit Cu-1 was obtained from the results obtained using a counter with a copper cathode [5]. The limit Cu-2 was set based on measurements made using the same counter at different temperatures (figure 3). The limit Al-1 was set based on data obtained by using a counter with an aluminum cathode presented at figure 4.

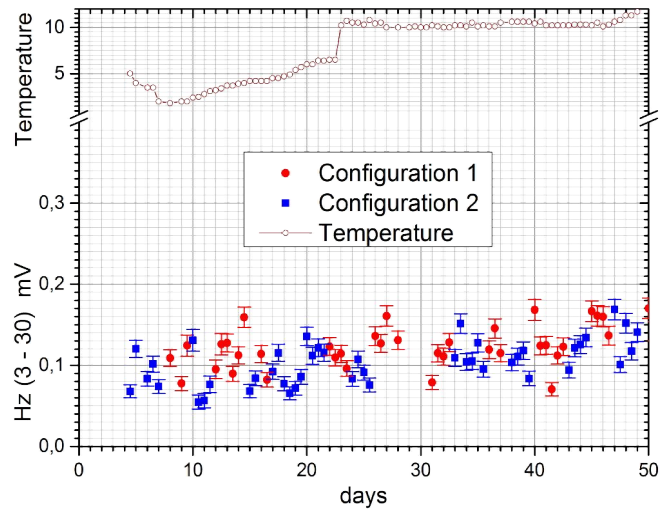

Figure 4. Count rates measured using a counter with an aluminum cathode.

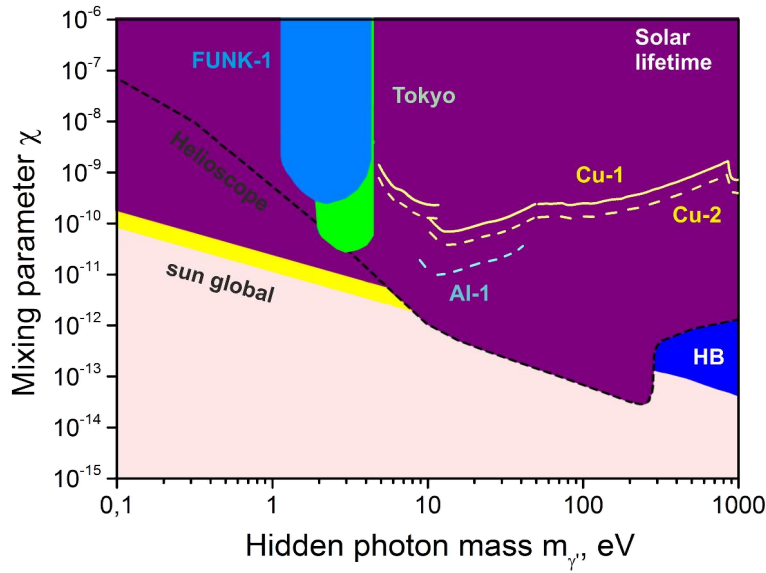

Figure 5. Limits obtained from a series of measurements Cu-1, Cu-2, and Al-1. Here limits Tokyo are from [3] and FUNK-1 from [7].

**4. Discussion.**

The limit for $\chi^2$ obtained by means of this technique depends critically upon the dark rate of the detector used for counting of single electrons. The dark rate can be reduced by applying more advanced technology of the surface treatment and by using lower temperatures of the detector to diminish the effect from thermal noise. We are planning also for further study to construct counters using materials with relatively high work function like nickel or platinum.

**Acknowledgments**

The authors acknowledge the financial support by Federal Agency of Scientific Organizations, Russia.